\documentclass[final,5p,times,twocolumn]{elsarticle} % Uncomment this to get the preview of how it will look once published.
\usepackage{amssymb}
\usepackage{lineno}
\usepackage{titlesec}
\titlespacing*{\section}
{0pt}{2mm}{0mm}
\titlespacing*{\subsection}
{0pt}{0mm}{0mm}

\usepackage[hidelinks]{hyperref}

\journal{Nuclear Instruments \& Methods in Physics Research, Section A}

\begin{document}

\begin{frontmatter}



\title{Characterization of timing and spacial resolution of novel TI-LGAD structures before and after irradiation}


\author[UZH]{M.~Senger}
\ead{matias.senger@cern.ch}

\author[FBK]{A.~Bisht}
\author[FBK]{G.~Borghi}
\author[FBK]{M.~Boscardin}
\author[FBK]{M.~Centis Vignali}
\author[FBK]{F.~Ficorella}
\author[FBK]{O.~Hammad Ali}
\author[UZH]{B.~Kilminster}
\author[UZH]{A.~Macchiolo}
\author[FBK]{G.~Paternoster}

\affiliation[UZH]{organization={University of Zurich Physic Institute},%Department and Organization
            addressline={Winterthurerstrasse 190}, 
            city={Zurich},
            postcode={CH-8057}, 
            state={Zurich},
            country={Switzerland}}

\affiliation[FBK]{organization={FBK-Sensors and Devices Micro Nano Facility},%Department and Organization
            addressline={Via Sommarive 18}, 
            city={Povo},
            postcode={I-38123}, 
            state={Trento},
            country={Italy}}

\begin{abstract}
The characterization of spacial and timing resolution of the novel Trench Isolated LGAD (TI-LGAD) technology is presented. This technology has been developed at FBK with the goal of achieving 4D pixels, where an accurate position resolution is combined in a single device with the precise timing determination for Minimum Ionizing Particles (MIPs). In the TI-LGAD technology, the pixelated LGAD pads are separated by physical trenches etched in the silicon. This technology can reduce the interpixel dead area, mitigating the fill factor problem. The TI-RD50 production studied in this work is the first one of pixelated TI-LGADs. The characterization was performed using a scanning TCT setup with an infrared laser and a $^{90}\textrm{Sr}$ source setup.
\end{abstract}


% \begin{highlights}
% \end{highlights}

% \begin{keyword}
% %% keywords here, in the form: keyword \sep keyword
% keyword one \sep keyword two
% %% PACS codes here, in the form: \PACS code \sep code
% %% MSC codes here, in the form: \MSC code \sep code
% %% or \MSC[2008] code \sep code (2000 is the default)
% \end{keyword}

\end{frontmatter}

\newcommand{\NEQ}{\textrm{~n}_\textrm{eq} \textrm{~cm}^{-2}}
\newcommand{\SRNINETY}{$^{90}\textrm{Sr }$}


\section{Introduction}
\label{sec:Introduction}
The TI-LGAD technology aims to produce a silicon detector for Minimum Ionizing Particles (MIPs) with spacial resolution in the order of $\lesssim 30 \textrm{~µm}$, temporal resolution of $\lesssim 30 \textrm{~ps}$ and withstaning high levels of radiation of the order of $\gtrsim 10^{15} \NEQ $ ~\cite{paternoster_2019_latests_developments, nicolo_VCI2022, paternoster_2021_novel_strategies}. In this technology a two-dimensional array of small pixels is implemented in the surface of a silicon die. Each pixel is an LGAD, i.e. a silicon detector with a gain implant providing a multiplication in the order of $\sim 5-20$. The isolation between neighbouring pixels is provided by physical trenches etched in the silicon~\cite{paternoster_2020_trench_isolated}.

\subsection{The TI-RD50 production}
\begin{figure}[h]
    \centering
    \includegraphics[width=.8\columnwidth]{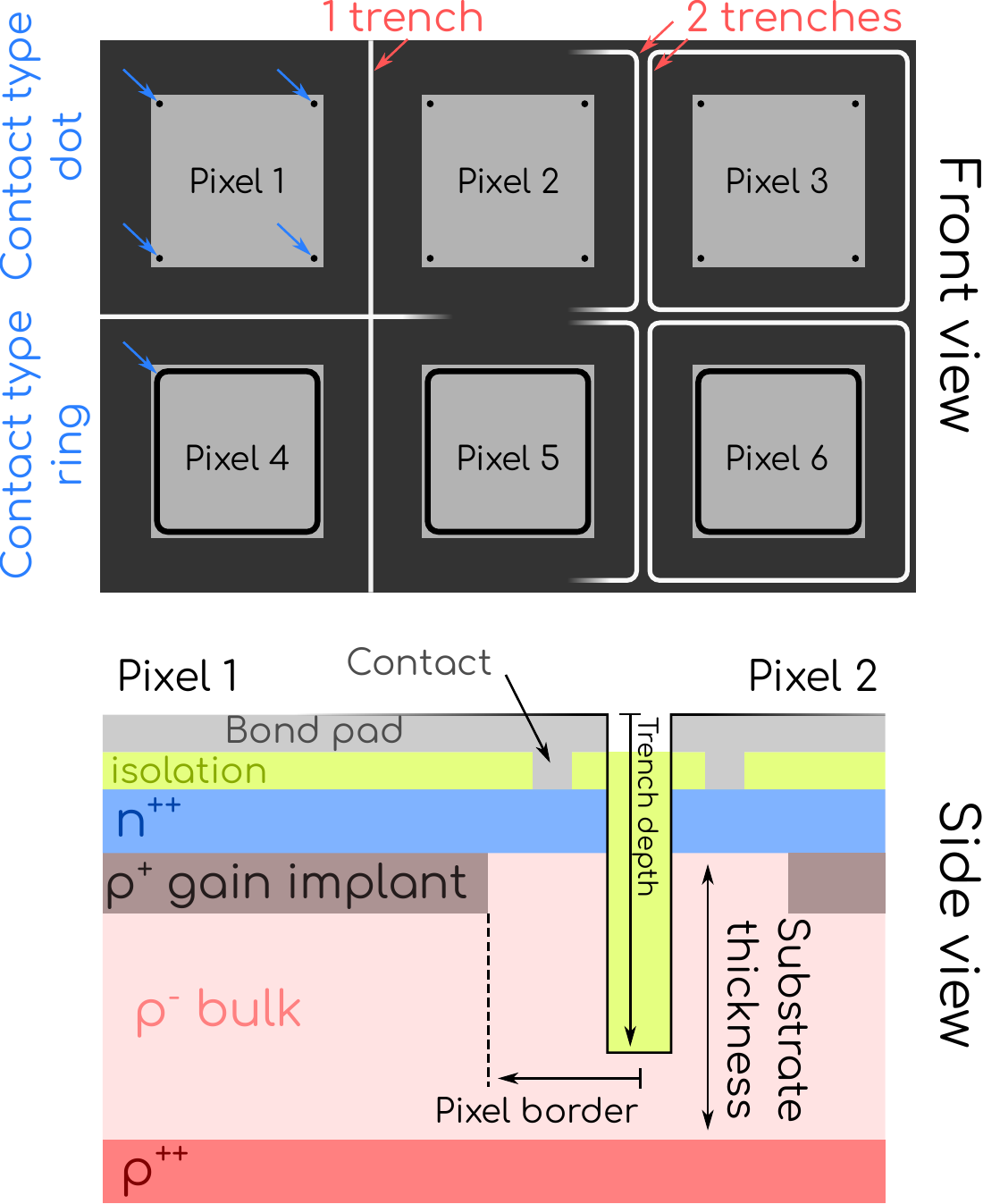}
    \caption{Schematic representation of the different design patterns in the \emph{TI-RD50} production. In the front view the difference between \emph{contact type dot} and \emph{contact type ring} and 1 or 2 trenches is shown. In the side view the \emph{trench depth} and \emph{pixel border} variables are ilustrated.}
    \label{Figure: schematic of design patterns}
\end{figure}

The \emph{TI-RD50} production, carried out by FBK within the RD50~Collaboration\footnote{\url{https://rd50.web.cern.ch/}} framework during 2020, implements several different designs with the objective of studying them and identifying the optimal ones. Figure~\ref{Figure: schematic of design patterns} illustrates each of the different design options. The first one is the \emph{contact type} which determines how the contact between the external metalization and the n+ implant is realized, either in a "dot" or "ring" shape. The \emph{number of trenches} separating each pixel can be either one or two. The \emph{trench depth} can take 3 values, "D1" (shallow), "D2" (intermediate) and "D3" (deep). Finally, the \emph{pixel border} is a measure of the distance between the trench and the gain implant and that can have 4 values "V1" (smallest, most aggressive), "V2", "V3" and "V4" (largest, most conservative). Devices were produced for all possible combinations of these variables. In this work, however, only devices with pixel border "V2" and "V3" were tested due to availability at the testing facility. All the devices studied in this work differ only by these variables, sharing all other design choices, such as the doses of each layer and the active thickness of $45 \textrm{~µm}$.

\subsection{Irradiation campaign}
As mentioned, the TI-LGAD devices are aimed to operate in high radiation environments. One possible application is a replacement of pixel disks in the CMS experiment during the Phase-3 upgrades. Here, irradiation levels up to fluences of 30$\times 10^{14} \NEQ$ are foreseen. Thus, we irradiated devices with reactor neutrons at JSI (Ljubljana) to three different fluences of $15$, $25$ and $35 \times 10^{14} \NEQ$. It has to be noted that the gain layer of most devices from the TI-RD50 production, and in particular those studied in this work, has not been optimized to withstand high fluences. In any case, the novelty here is the introduction of the trenches as the isolation mechanism between neighbouring pixels, and our study focuses on this. Radiation hard gain layer designs and carbon coimplantation should be compatible with the trench isolated pixels in future productions.

\section{Setup and measurements}
The measurements were performed at the laboratory for silicon devices of the CMS group at UZH. A \emph{scanning Transient Current Technique} (TCT) setup was used. This setup has an infrared laser (1064~nm) with a Gaussian profile with $\sigma \approx 9 \textrm{~µm}$. The laser can be focused in arbitrary positions on the sample with a precision better than $1 \textrm{~µm}$ thus allowing to study the properties of the Device Under Test (DUT) as a function of position. In addition, the laser pulses are split in two paths, one containing 20 extra meters of optic fiber, before being sent to the DUT. This produces two almost identical pulses of light with a fixed temporal separation of 100~ns and allows to study the time resolution of the DUT. More details and pictures can be found in~\cite{senger_2022_VCI_presentation}.

\begin{figure}[h]
    \centering
    \includegraphics[width=.8\columnwidth]{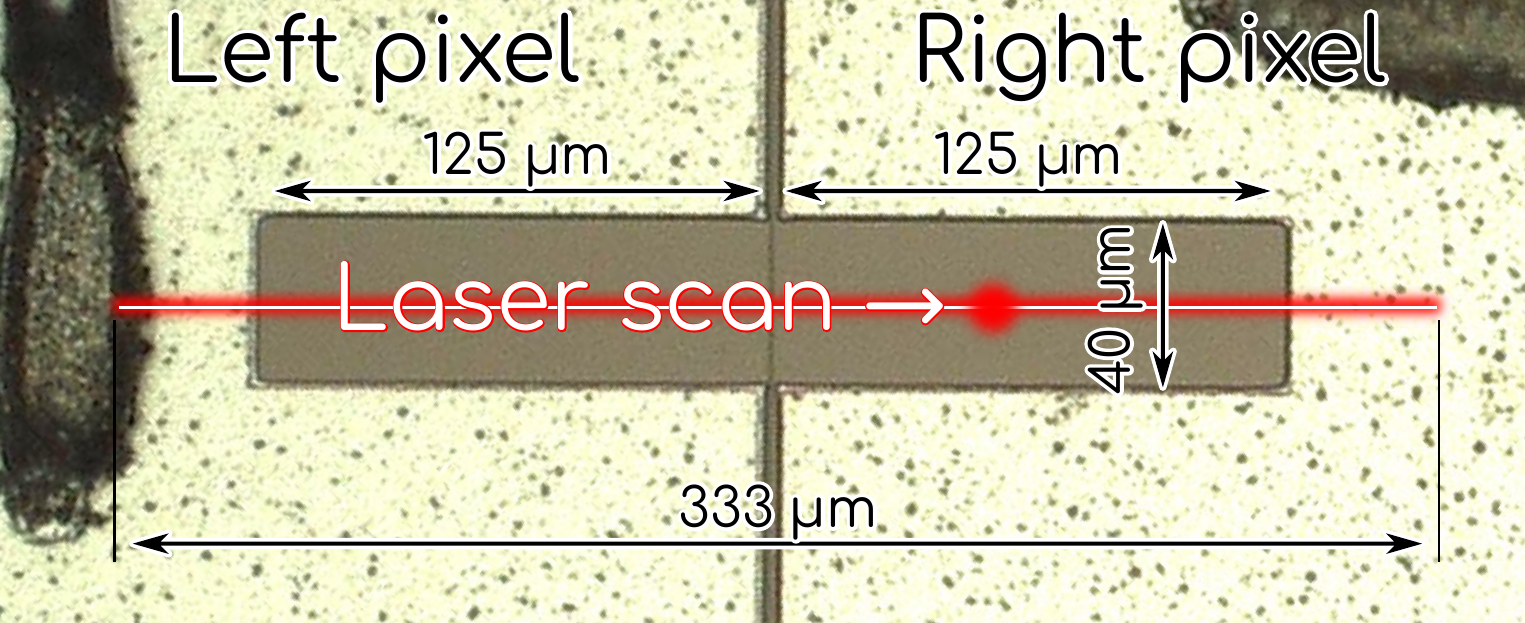}
    \caption{Microscope picture of one device showing the area between two pixels and the opening in the metalization through which the laser was shone into the silicon. The thin dark line that can be seen vertically dividing each pixel is one trench.}
    \label{Figure: picture of window with laser}
\end{figure}

The samples were installed in a custom made passive readout board. This board had incorporated a temperature and humidity sensor very close to the DUT, which was used to stabilize the temperature at -20~°C during the measurement of irradiated devices. The samples had either 4 or 16 pixels, but only two of them were used while the others were connected to ground. In particular, pixels with a special opening in the metalization were measured, which allow the light from the laser to reach the silicon, as shown in Figure~\ref{Figure: picture of window with laser}.

Each pixel was connected to a broadband fast amplifier (Cividec C2HV) and then into the oscilloscope, which was either a LeCroy WaveRunner 640Zi or 9254M. The samples were biased using a Keithley 2470 high voltage power supply, which was also used to trace the IV curves. 

An example of the charge collected as a function of position is shown in Figure~\ref{Figure: pixel isolation}. The inter-pixel distance was defined as the distance between the two positions where 50~\% of the maximum charge is collected in the left and right pixels, which is a criterion that was also used in the past in other works~\cite{siviero_2019_lab_measurements_of_TI_LGAD,ferrero_et_al_2021_book}.

The time resolution was calculated at each position by shining the laser many times and then applying the constant fraction discriminator method~\cite{ferrero_et_al_2021_book} between each of the two pulses produced after the laser splitting system described before. More details in reference~\cite{senger_2022_VCI_presentation}.

The laser intensity for all the measurements was configured such that the mean value of the collected charge is close to the most probable value of the distribution obtained with beta particles from a radioactive \SRNINETY source. This calibration was done prior to the measurements.

\section{Results}
For each value of \emph{contact type}, \emph{number of trenches}, \emph{trench depth} and \emph{pixel border} a device was measured, and the same method and data analysis were applied to all the samples. In the following subsections some of the most relevant measured parameters will be presented.

\subsection{Pixel isolation provided by the trenches}
\begin{figure}[h]
    \centering
    \includegraphics[width=1\columnwidth]{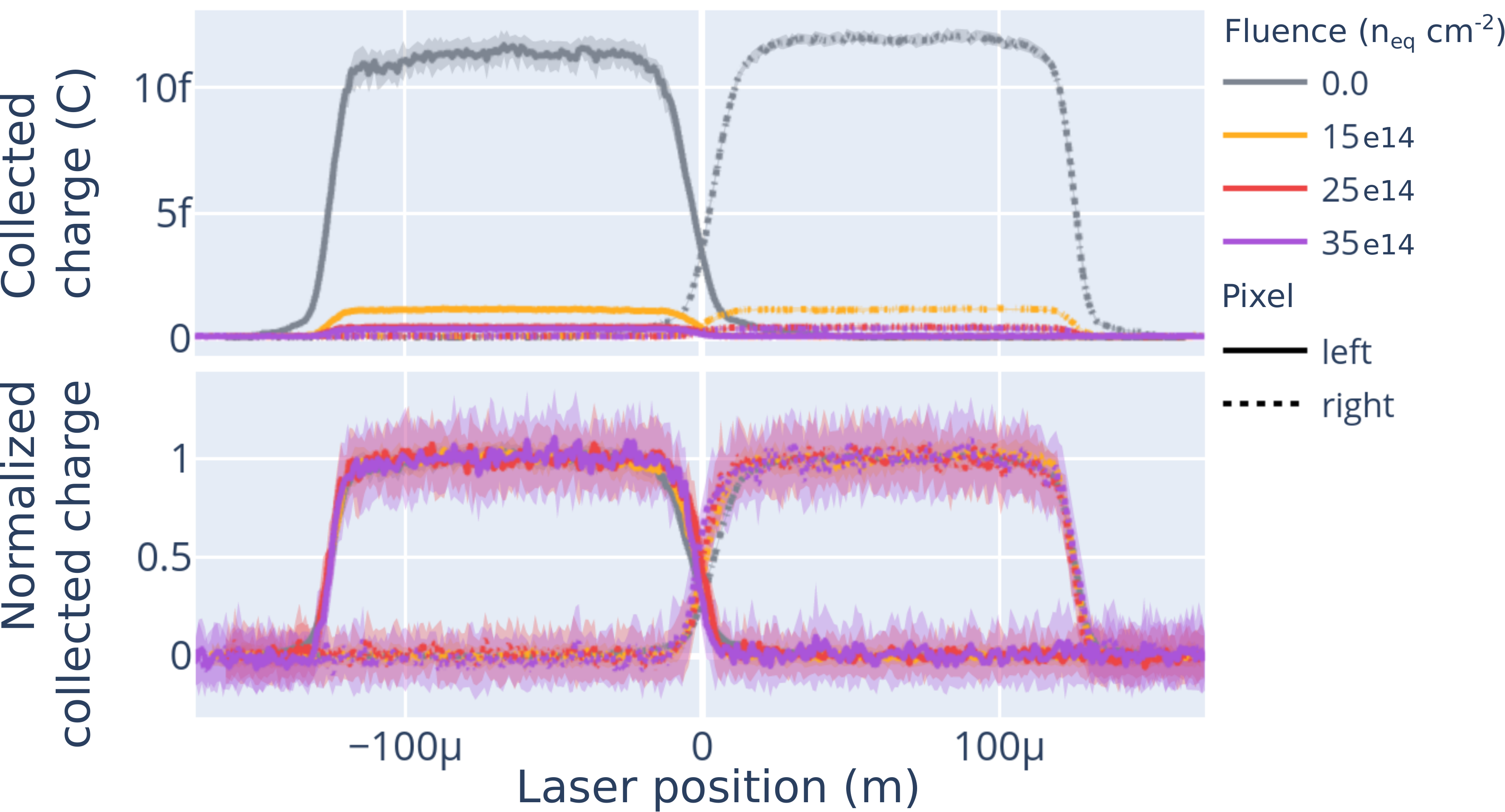}
    \caption{Collected charge and normalized collected charge as a function of position for four identical devices (contact "dot", 1 trench, depth "D1", border "V2") irradiated to different fluences. Each trace was normalized individually between 0 and~1.}
    \label{Figure: pixel isolation}
\end{figure}

The results in Figure~\ref{Figure: pixel isolation} show an excellent pixel isolation for all the fluences studied as no cross-talk could be observed in any case. This demonstrates that the isolation provided by the trenches is very good and withstands without issues fluences of at least $35 \times 10^{14} \NEQ$. 

\begin{figure}[h]
    \centering
    \includegraphics[width=1\columnwidth]{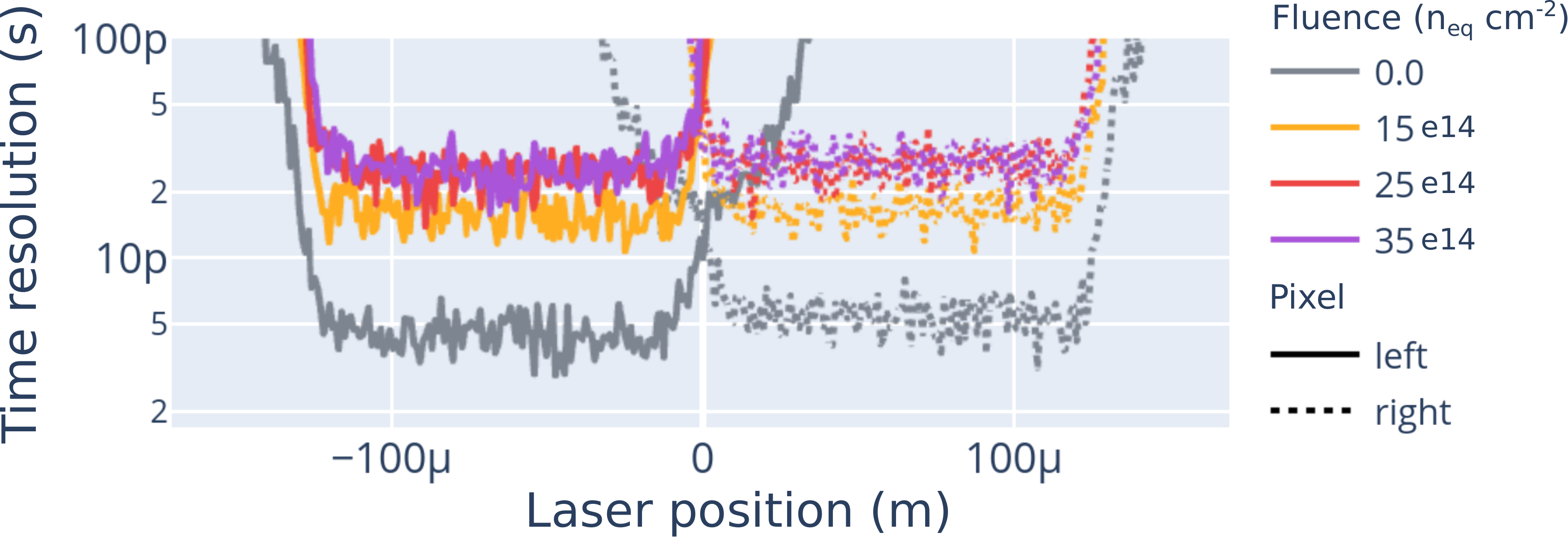}
    \caption{Time resolution as measured with the TCT as a function of position for the same four devices shown in Figure~\ref{Figure: pixel isolation}.}
    \label{Figure: time resolution vs position}
\end{figure}

\subsection{Gain}
\begin{figure}[h]
    \centering
    \includegraphics[width=1\columnwidth]{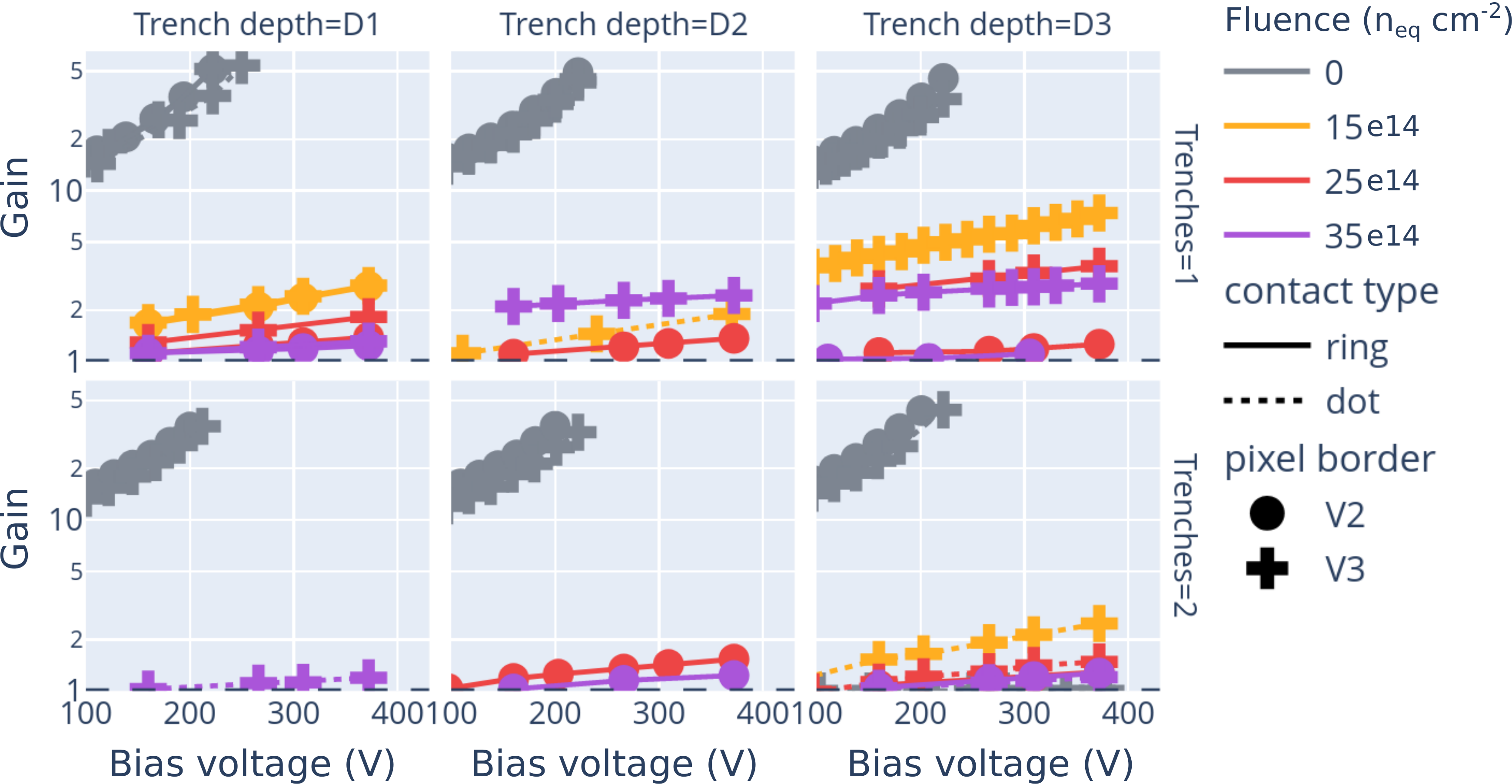}
    \caption{Measured gain vs bias voltage for each of the different design patterns and fluences studied.}
    \label{Figure: gain vs bias voltage}
\end{figure}

For each of the devices studied the gain was computed as the ratio between the collected charge by the DUT and the collected charge by a non irradiated PIN diode\footnote{I.e. a device with no gain layer.} of the TI-RD50 production. The results are shown in Figure~\ref{Figure: gain vs bias voltage} as a function of the bias voltage. We see that before irradiation the gain reaches values between 30 and 50 at the operating voltage of 200~V while after irradiation it is severely degraded to values between 3 and 8 for the lowest fluence and between 1.2 and 3 for the highest fluence. It has to be noted that for the irradiated devices the gain could not be measured at higher voltages than 400~V as to the reference PIN diode went into breakdown. At 600~V a slightly higher gain was observed as the collected charge was larger~\cite{senger_2022_VCI_presentation}, anyhow its precise value could not be determined due to the breakdown of the reference PIN diode. We don't observe any systematic dependence of the gain on each of the design patterns studied.

\subsection{Inter-pixel distance}
\begin{figure*}[h]
    \centering
    \includegraphics[width=.8\textwidth]{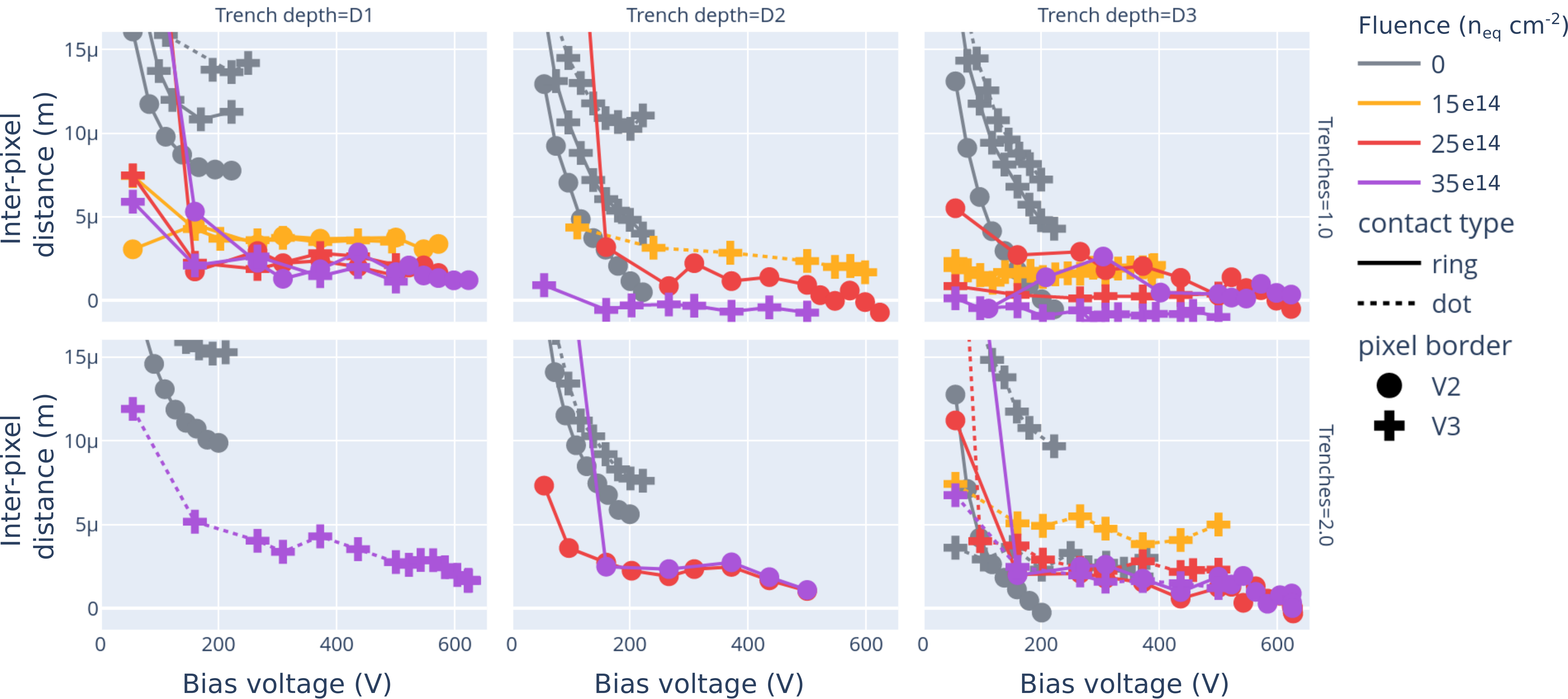}
    \caption{Measured inter-pixel distance as a function of the bias voltage for each of the different design patterns.}
    \label{Figure: ipd vs bias voltage}
\end{figure*}

The inter-pixel distance (IPD) is one of the most relevant properties analyzed in this study. It was measured for each design pattern at each different irradiation fluence for different values of the bias voltage. The results are shown in Figure~\ref{Figure: ipd vs bias voltage}. It can be seen that before irradiation this quantity has an important dependence with the bias voltage as well as with the different design variables. We note that the designs with Pixel border V2 achieve lower IPD values than V3, as well as deeper trenches with respect to shallow ones, single trenches with respect to double trenches, and contact type ring with respect to dot.  After irradiation all these dependencies on the design features are much less relevant. We still observe a small improvement with increasing bias voltage. In a previous study on non irradiated TI-LGADs the IPD was also seen to improve by increasing the bias voltage~\cite{bisht_2021_TILGAD_at_zurich_workshop}. 

\subsection{Time resolution}
\begin{figure*}[h]
    \centering
    \includegraphics[width=.8\textwidth]{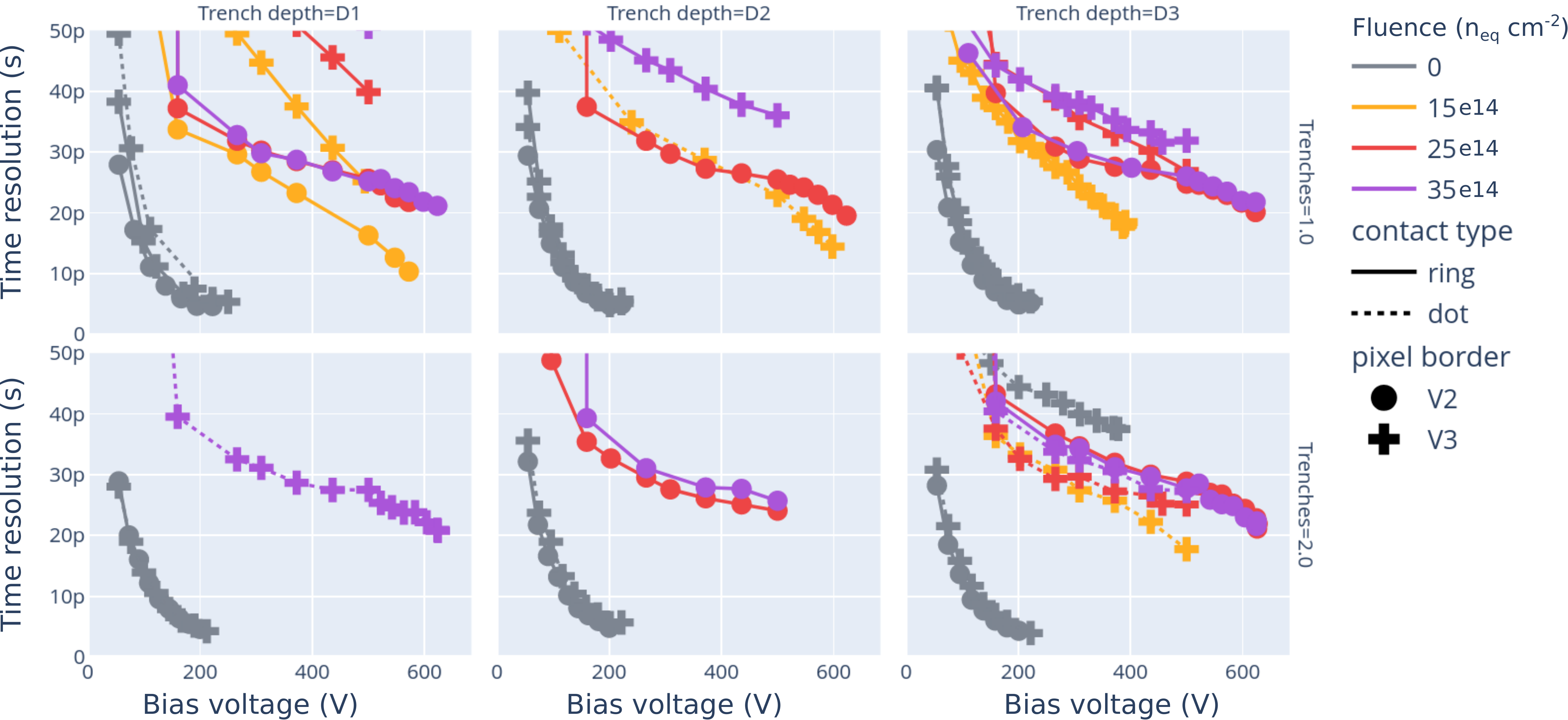}
    \caption{Measured time resolution (in the TCT) as a function of the bias voltage for each of the different design patterns.}
    \label{Figure: time resolution vs bias voltage}
\end{figure*}

The time resolution is another important quantity that was measured for the TI-RD50 production. The results are shown in Figure~\ref{Figure: time resolution vs bias voltage}. As previously discussed the time resolution was determined in the TCT setup using a constant fraction discriminator method. The first observation is that prior to irradiation all devices show roughly the same time resolution, achieving a value of about 5~ps at 200~V. The time resolution shows no dependence with the different design patterns studied. After irradiation we see that the time resolution is degraded, however it is still good. All the devices measured up to 600~V show time a resolution on the order of 20~ps. 

As can be observed in Figure~\ref{Figure: time resolution vs position} the time resolution is practically uniform along the pixel up to very close to the inter-pixel area. This is an important characteristic as it makes the time resolution independent of the impact position.

\section{Measurements with beta source}
In order to investigate the contribution of the Landau term to the time resolution of the TI-LGADs~\cite{ferrero_et_al_2021_book}, a subset of the irradiated devices was measured with a radioactive \SRNINETY beta source setup. These measurements are essential to give a measurement of the timing resolution that is comparable with the expected performance of the pixelated LGAD devices in a real HEP detector environment. The same constant fraction discriminator analysis was performed with one of the pulses being that from the DUT while the other being that of a reference detector, which was a previously calibrated single pad LGAD device. The electrical connections for the DUT were kept completely the same as in the TCT setup. In this case time resolution between 50 and 65~ps was observed for devices irradiated at $35 \times 10^{14} \NEQ$.

\section{Annealing effects}
Three devices, all with the same design but with different values of fluence, were measured in the TCT after irradiation and then left to anneal for 7 days at room temperature. After this annealing period the TCT measurements were repeated and compared with those before annealing. The effects of the annealing were observed to behave systematically in the three detectors. Such effects demonstrated no improvement in their performance, with the exception of the bias current which was reduced approximately by a factor of 40~\%. The IPD was unaffected by the annealing process while the gain was reduced by a factor between 10 and 20~\% and the time resolution worsened by a factor between 15 to 20~\%. 

\section{Conclusions}
The first production of trench isolated LGAD pixelated devices, TI-RD50, was extensively characterized both before and after being irradiated with neutrons up to fluence values of $35\times 10^{14} \NEQ$. The inter-pixel distance of many of the different designs in this production was studied using a scanning TCT setup. It was found that not all the designs behave the same and the that the best combination among those studied is one trench, deep, with pixel border V2 and contact type ring. It was observed that after irradiation the inter-pixel distance becomes less dependent on the bias voltage as well as on the different design patterns. 

The pixel isolation provided by the trenches was observed to be very good both before and after irradiation and for all the different designs. The cross-talk levels were in all cases below the noise floor of the measurement setup.

The time resolution was studied. Values on the order of 5~ps were observed before irradiation while values on the order of 20~ps were observed after irradiation, both using infrared laser in the TCT setup. As opposed to the inter-pixel distance, the time resolution did not show a systematic dependence on the different design patterns. In order to investigate the expected timing performance with MIPs, for a few irradiated samples the time resolution was also measured with a beta source and values on the order of 50-65~ps were observed.

From these results we conclude that the TI-LGAD technology is a promising candidate for the implementation of the 4D-pixels as they can provide good time and space resolution with reasonable fill factor while withstanding high radiation levels.

\section*{Acknowledgements}

This project was conducted within the RD50 Collaboration and has received 
funding from the European Union’s Horizon 2020 Research and Innovation program under grant agreement No~101004761. 

\appendix

\end{document}